\documentclass[12pt, draftclsnofoot, onecolumn]{IEEEtran}
\usepackage{cite}
\usepackage{amsmath,amssymb,amsfonts}
\usepackage{algorithm}
\usepackage{algorithmic}
\usepackage{graphicx}
\usepackage{textcomp}
\usepackage{xcolor}
\usepackage{multicol}
\usepackage{subcaption}
\usepackage{mathrsfs}
\usepackage{soul}
\usepackage{setspace}
\usepackage{listings}

\begin{document}
\onecolumn

\title{Phase Spectrometry For High Precision mm-Wave DoA Estimation In 5G Systems}

\vspace{5mm}
\author{\IEEEauthorblockN{Farzam Hejazi, }\textit{Department of Electrical and Computer Engineering}\\
\textit{University of Central Florida}\\
Orlando, USA, 
farzam.hejazi@ucf.edu\\
\and
\IEEEauthorblockN{ Nazanin Rahnavard,}{\textit{ Department of Electrical} \textit{and Computer Engineering,} 
\textit{University of Central Florida}\\
Orlando, USA,
nazanin@eecs.ucf.edu}
}

\maketitle
\begin{abstract}

In this paper, we introduce a direction of arrival (DoA) estimation method based on a technique named phase spectrometry (PS) that is mainly suitable for mm-Wave and Tera-hertz applications as an alternative for DoA estimation using antenna arrays. PS is a conventional technique in optics to measure phase difference between two waves at different frequencies of the spectrum. Here we adapt PS for the same purpose in the radio frequency band. We show that we can emulate a large array exploiting only two antennas. To this end, we measure phase difference between the two antennas for different frequencies using PS. Consequently, we demonstrate that we can radically reduce the complexity of the receiver required for DoA estimation employing PS. We consider two different schemes for implementation of PS: via a long wave-guide and frequency code-book. We show that using a frequency code-book, higher processing gain can be achieved. Moreover, we introduce three PS architectures: for device to device DoA estimation, for base-station in uplink scenario and an ultra-fast DoA estimation technique mainly for radar and aerial and satellite communications. Simulation and analytical results show that, PS is capable of detecting and discriminating between multiple incoming signals with different DoAs. Moreover, our results also show that, the angular resolution of PS depends on the distance between the two antennas and the band-width of the frequency code-book. Finally, the performance of PS is compared with a uniform linear array (ULA) and it is shown that PS can perform the same, with a much less complex receiver, and without the prerequisite of spatial search for DoA estimation. 

\end{abstract}
\section{Introduction}
5G mobile networks promise to bring a new era of ultra high-speed communications that surpasses previous generations by several order of magnitudes in communication capacity \cite{boccardi2014five}. One of the core technologies behind such a spectacular revolution is spatial devision multiple access (SDMA). SDMA enables massive Multi-Input-Multi-Output (MIMO) communication by providing an ability to focus energy on users' devices, empowering pushing the capacity of the network to such a immense boundaries required for 5G communications \cite{agiwal2016next,hejazi2021dyloc}. Simultaneously, mobile mm-wave communication is enabled through 5G networks, that transform directional communication from a promising aspect of next generation networks, into a must-have feature \cite{agiwal2016next,rappaport2013millimeter}. Mm-wave communication experiences huge attenuation in the open air, therefore the transmitted energy needs to be directed into narrow rays, to meet sufficient signal-to-noise-ratio (SNR) thresholds required at receivers \cite{rappaport2019wireless}. In addition to 5G applications, DoA estimation is a required aspect of UAV-to-device and satellite-to-device high frequency and ultra high-speed communication \cite{agrawal20165g}. Moreover, mm-wave and Terahertz radars used for autonomous driving exploit DoA estimation techniques to estimate angles of the objects around \cite{dickmann2016automotive}. As directional communication has gained importance in new generation communications, DoA estimation has obtained gravity as an enabler of directional communication. To clarify this necessity, consider that any two devices that exploit directional antennas cannot communicate unless they ascertain in which direction they should send/receive signals to/from the other device. Moreover, this knowledge of angle (or position) of the other device should be maintained during the communication period otherwise the link will be disrupted \cite{nitsche2014ieee}. 

The most common DoA estimation techniques use directional antennas mounted on both the transmitter and the receiver to obtain the initial guess of the relative angle between two devices, this process is also referred as initial access (IA) \cite{giordani2016initial}. To fulfil this strategy, the first device starts searching for the second device through a beam training protocol, until it finds the other device. Next, the second device repeats the same procedure until the link is established, at this point, they employ tracking techniques to maintain the directional connection between them \cite{giordani2018tutorial}.
Although such a strategy looks favorable for DoA estimation, it is highly probable that it does not work well when a large number of devices are packed into a specific area or in the presence of a strong multi-path between two device. Moreover, beams are most of the time busy with beam training/tracking searches instead of transmission/reception which reduces the communication capacity \cite{giordani2019standalone}. In other words, using the same antenna for communication and direction-finding, requires using a common resource for two inherently antithetical task in terms of directional antenna requirements. Higher communication capacity requires highly directional antennas to reduce interference and to maximize signal power at the receiver, conversely, as antenna's beams become narrower the beam training/tracking periods increase and consequently the overhead escalates which eventually reduces the effective communication capacity. To overcome deficiencies of such a strategy we propose to avoid using directional antennas for DoA estimation at both sides of the link, and estimate DoA based on measuring phase difference of arrival (PDoA) of signal between two antennas mounted on the device for multiple frequencies. Meanwhile, we can allocate a directional antenna exclusively for communication purposes. In our proposed strategy, we avoid spatial search to establish the link in the first place, on the other hand, we rely on the received signal in two omni-directional antennas. Subsequently, we amplify the attenuated received signal by a huge processing gain, then estimate DoAs of all of propagation paths between two devices.
We will show that exploiting our proposed technique, we can convert spatial search duration to a means to increase DoA estimation precision, and more importantly, we can allocate a specific highly directional antenna for communication, and consequently take advantage of the whole communication capacity such a directionality provides.

In our proposed technique, two antennas are mounted on the device with several mm gap between them, and the PDoA of signal measured through a novel technique named standing wave spectrometry for multiple frequencies. Standnig wave spectrometry is widely used in optical applications to measure phase difference between two rays at multiple frequecies of the optical spectrum \cite{sabry2015monolithic}\cite{wolffenbuttel2005mems}\cite{jovanov2010standing}. To the best of authors knowledge, it is the first time that this technique is introduced for RF mm-wave applications. By applying spectrometry not only we can estimate the DoA of a signal precisely, but also we can estimate multi-path DoAs and the power of each path for a mm-wave propagation environment. Although the proposed approach is inherently a wide-band (WB) technique, it does not require ultra high speed sampling rates essential for must of WB techniques. Consequently, the proposed technique provides us with two main advantages: more data about the DoA of incoming signal, and reduced cost and complexity of the receiver. The first is obtained by discriminating between all incoming propagation paths between the source and the device. The second is secured by greatly reducing the complexity of the DoA estimation through simplification of the receiver by bypassing signal down-conversion and reducing the number of required antennas. Furthermore, we show that the proposed phase difference measurements equals to highly accurate measurement of time difference of arrival (TDoA) of signal between two antennas in the Fisher sense. Moreover, we will prove that the cramer-rao lower bound of error (CRLB) of DoA estimation using the proposed technique equals to a uniform linear array (ULA) that employs multiple antennas, in the Fisher sense.


\section{Related Works}

DoA estimation techniques have plethora of applications in Radar, Sonar and Electronic Ware-fare (EW) literature. In these applications, DoA estimation is mainly used to find the relative direction between two objects. Primitive DoA estimation techniques use pencil beam antennas (e.g. dish antennas) along with mechanical actuators for steering the beam and spatial search \cite{skolnik2001radar},\cite{barshan1992bat},\cite{poisel2012electronic},\cite{hejazi2013lower},\cite{hejazi2013new},\cite{khalili2013secant}. More recent techniques, use beamforming techniques over array antennas to obtain narrow beams. In beamforming, input/output of each antenna of an array, is multiplied by a weight (e.g. a phase shift) to form a desired beam shape. In beamforming, there is no need for mechanical steering, and beams can be steered electronically by changing weights of the antennas. Spatial scanning provided by beamforming proves to be much more faster than the mechanical scanning, moreover, can generate multiple beams simultaneously. Therefore, modern phased array radars can search the environment very fast, and can track and engage with multiple targets concurrently \cite{mailloux2017phased}.

Recently, DoA estimation also has gained attention as an enabler of ultra-high-speed (Multiple Gbps) directional communications between two devices or a base-station and multiple devices. 5G communication mainly utilizes advanced beamforming capabilities and array antennas for directional communications. 3 different architectures has been introduced for beamforing for 5G applications: 1-Analogue 2-Digital 3-Hybrid \cite{kutty2015beamforming}. In Analogue beamforming, the beam is shaped via a single RF chain, and so only one beam can be shaped in each time slot. This structure is more power efficient compared to the two other architectures, however, is not as flexible as them in generating multiple beams. Digital beamforing, allocates a specific Rf chain and data-convertor for each antenna and potentially can generate several beams simultaneously. This structure is the most flexible one, however is very power hungry and complicated in comparison to other techniques \cite{yang2018digital}. Hybrid beaforming scheme assigns multiple RF chains for antennas, while, the number of RF chains is less than the number of antennas. This type of beamforming is the most common scheme for 5G applications, since it can balance a trade-off between complexity, flexibility and power consumption \cite{sohrabi2016hybrid,molisch2017hybrid}. All directional antennas powered by various beamforming architectures require spatial search to initiate a communication link . Giordani et. al showed that overhead caused by beam-training protocols heavily limits number of array elements at both base stations and user equipments, moreover, several milliseconds is required to establish a link between a base station and user equipment \cite{lien20175g}. 

Interferometric wide-band DoA estimation, has been widely investigated in EW and lightning localization applications \cite{mardiana2000broadband},\cite{wu1995direction}, \cite{hejazikookamari2018novel},\cite{kookamari2017using},\cite{hejazi2014sar},\cite{hejazi2020tensor},\cite{hejaziwireless},\cite{joneidi2019large}. In this technique PDoA of signal between two antennas placed more than half-wavelength apart is measured. Since the phase difference is ambiguous and can represent several DoAs, a number of techniques has been introduced to disambiguate the phase. These techniques include: correlative interferometry (CORR), second order difference array (SODA), SODA-Base Inference (SBI) and Common Angle Search (CAS). CORR employs PDoAs between at least two pairs of antennas and compare measurements with a pre-prepaired database of measurements to determine DoA \cite{kebeli2011extended}. SODA and SBI operate an additional antenna pair with less than half a wavelength gap between antennas to  translate PDoA to an unambigeous DoA. SODA and SBI only works well when input SNR is high enough \cite{mollai2018compact,mollai2019wideband}. CAS utilizes two or more antenna pairs and introduces the common angle recommanded by all PDoAs as the unambiguous DoA \cite{searle2017disambiguation}. These techniques can estimate DoA very precisely in a wide-band frequency range, however, none of them can distinguish between DoAs, if two or more signals with differnet DoAs are received simultaneously at the antenna pairs. 

Here in section \ref{Fisher}, we prove that phase interferometry meaurements (PIM) between two antennas equals to highly precise time difference of arrival (TDoA) measurements in the Fisher sense. Moreover, we demonstrate that DoA estimation using PIM between two antennas several wavelength apart equals to DoA estimation using a large ULA in the Fisher sense. Since PIMs represent ambigeous DoAs, we introduce phase spectrometry (PS) to disambiguate PDoAs in section \ref{PS}. In contrast with Interferometric DoA estimation, we prove that PS can distinguish between multiple concurrent DoAs. Furthurmore, we introduce standing wave receiver (SWR) to extract PDoAs, which is much less complicated than beamforming receivers. We explain how SWR does not need any down-conversion or high sampling rates to extract PDoA. In section \ref{DoARes} we investigate DoA estimation resolution provided by PS. Then we introduce two approaches to implement PS, one through a long wave-guide, another via employing a frequency code-book in section \ref{longSW} and \ref{FreqCB} respectively. In section \ref{SNR}, we analyse SNR improvement caused by PS. Furthermore, we will show how the whole time required by directional techniques for spatial search can be effectively consumed in PS to improve DoA estimation precision.  We discuss the ability of the proposed technique to identify  DoA of signals from several devices in both uplink and downlink scenarios in section \ref{scale}. Moreover, we introduce an alternative architecture of the technique that provides us with ultra-fast DoA estimation capability in section \ref{UFast}. In Section \ref{sim}, we examine PS performance via various simulations. Finally we conclude the paper in section \ref{conc}. 

\section{Phase Interferometry Measurements}
\label{Fisher}
\begin{figure}
    \centering
    \includegraphics[width=3in,height=2.3in]{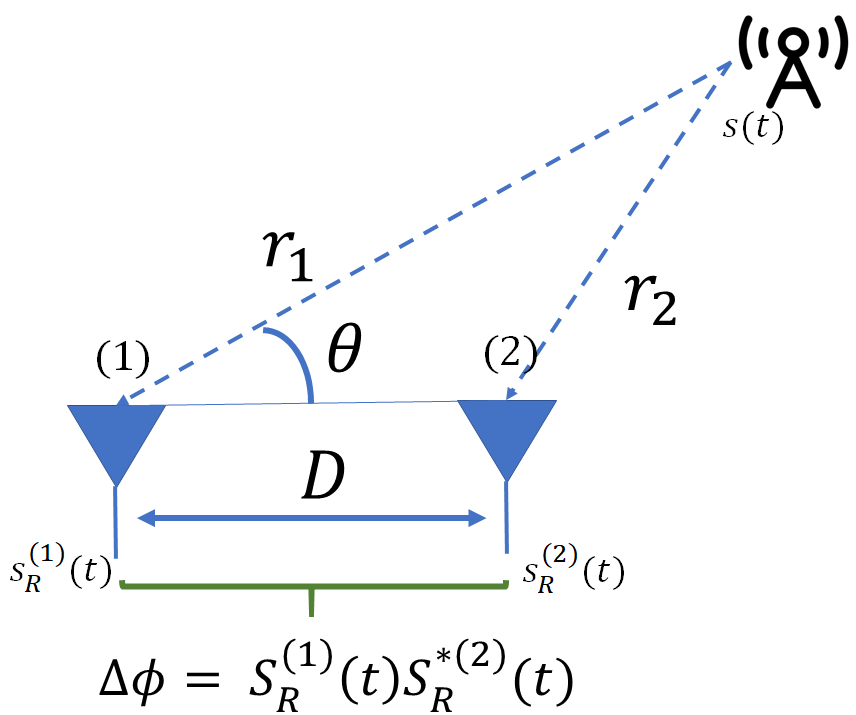}
    \caption{PIM illustration, two antennas implemented on a device receive a signal ($s(t)$) emitted by a source ($s^{(1)}_{R},s^{(2)}_{R}$). PIM is defind as the interaction of two signals $\Delta \phi = s^{(1)}_{R}s^{*(2)}_{R} $}
    \label{PIMDEFGEO}
\end{figure}

Consider 2 antennas with gap $D$ mounted on a device (Figure \ref{PIMDEFGEO}), referred as phase interferometry array (PIA), both of them are receiving a signal emitted by a source $s(t)$.  The signal is a monotone with carrier frequency $f_c$   
\begin{equation}
    s(t)=a \: e^{j2\pi f_c t}\,,
    \label{sigmod}
\end{equation}
where $a$ is the amplitude of the signal. Both the first and the second antennas receive the signal, denoted by $s^{(1)}_{R}(t)$ and $s^{(2)}_{R}(t)$ respectively,  with a relative delay $\Delta (t)$ which results in a phase difference between two signals. We define phase interferometry measurements (PIM) as 
\begin{equation}
    \Delta \phi = s^{(1)}_{R}(t)s^{*(2)}_{R}(t)=a^2_R \: e^{j2\pi f_c \Delta (t)}+v_n=b e^{j2\pi f_c \Delta (t)}+v_n  \,,
\end{equation}
where $a_R$ is the amplitude of the signal received at the PIA and $v_n$ is white noise, we also refer to $e^{j2\pi f_c \Delta (t)}$ as PDoA throughout this paper. In the next section we prove that PIM is equivalent to DoA estimation using a ULA in the Fisher sense. 
\subsection{Fisher Information Matrix of PIM, TDoA \& DoA}
\begin{figure}
    \centering
    \includegraphics[width=4.9in,height=3in]{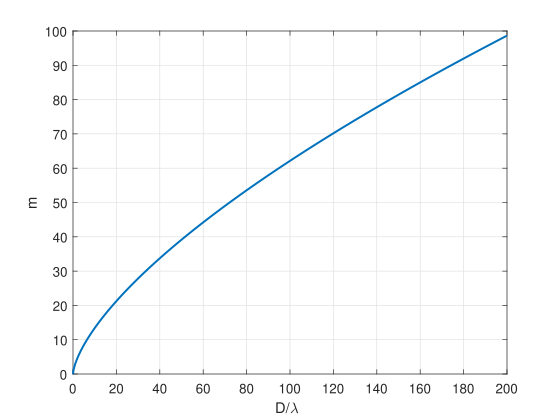}
    \caption{$m$ versus $\frac{D}{\lambda}$, where $m$ is the number of array elements of a ULA that is equivalent to (in the Fisher sense) a phase interferomery array (PIA) with gap $D$ between two antennas}
    \label{mdlambda}
\end{figure}

Given noise is Gaussian and independent for each PIM, Fisher information matrix (FIM) of $\Delta \phi $ with respect to an arbitrary vector $\boldsymbol{x}$ , e.g. unknowns to be estimated, can be derived as \cite{farina1999target}
\begin{align}
\sum_{\mathbb{P}} \frac{1}{\sigma^2} \nabla_{\boldsymbol{x}} \Delta \phi^{H} \nabla_{\boldsymbol{x}} \Delta \phi=&\sum_{\mathbb{P}} \frac{1}{\sigma^2} (-j2\pi f_c \nabla_{\boldsymbol{x}} (\Delta (t))^{H} be^{-j2\pi f_c \Delta (t)})  (j2\pi f_c \nabla_{\boldsymbol{x}} \Delta (t) be^{j2\pi f_c \Delta (t)}))= \nonumber \\
&\sum_{\mathbb{P}} \frac{4b^2\pi^2 f^2_c}{\sigma^2} \nabla_{\boldsymbol{x}} \Delta (t)^{H} \nabla_{\boldsymbol{x}} \Delta (t) \,,
\end{align}
where $\mathbb{P}$ is the set of all PIMs, and $\nabla_{\boldsymbol{x}}$ is the gradient operator with respect to (w.r.t) $x$. Therefore, FIM of PIM is exactly equals to the following observations, 
\begin{equation}
\delta (t)=b\Delta (t)+\frac{v_s}{2\pi f_c} \,.
\label{delti}
\end{equation}
where $\delta (t)$ is an observation of TDoA of signal between two antennas. Therefore, PIM with additive white noise power $\sigma^2$ equals to TDoA observations with additive white noise power $\frac{\sigma^2}{4\pi^2 f^2_c}$ of the same PIA in the fisher sense. Assuming far field criteria is fulfilled \cite{chen2002source}, we have
\begin{equation}
\delta (t)=b\frac{D}{c} cos(\theta_A)+\frac{v_s}{2\pi f_c} \,,
\label{delti1}
\end{equation}
where $c$ is the speed of light and $\theta_A$ is DoA of signal and $D$ is the gap between two antennas. CRLB of $\theta_A$ estimation based on measurements as of (\ref{delti1}) can be derived as follows
\begin{equation}
\mathrm{CRLB}_{\theta_{A}}=\frac{\frac{\sigma^2_s}{b^2}}{(\frac{D}{c}2\pi f_c)^2 sin^2(\theta_A)}=\frac{\frac{\sigma^2_s}{b^2}}{(\frac{2\pi D}{\lambda})^2 sin^2(\theta_A)}\,.
\label{CRMPIM}
\end{equation}
Now lets take a look at CRLB of DoA estimation using a ULA in which antennas are placed half wavelength apart \cite{penna2011bounds},
\begin{equation}
\mathrm{CRLB}_{\theta_{A}}=\frac{6 \frac{\sigma^2_s}{b^2}}{\pi^2 m(m^2-1)sin^2(\theta_A)} \approx \frac{6 \frac{\sigma^2_s}{b^2}}{\pi^2 m^3 sin^2(\theta_A)}\,,
\label{CRBULZ}
\end{equation}
where $m$ is the number of array elements. Given the same SNR, DoA estimation using PIM and a ULA array are equivalent in the Fisher sense when, 
\begin{equation}
m=(24)^{\frac{1}{3}}(\frac{D}{\lambda})^{\frac{2}{3}} \approx 2.8845 (\frac{D}{\lambda})^{\frac{2}{3}} \,.
\label{m/d}
\end{equation}
Figure \ref{mdlambda} illustrates \eqref{m/d}, as an example, DoA estimation using a PIA with $\frac{D}{\lambda}=200$ is equivalent to a ULA with 100 elements in the Fisher sense.  Consequently, DoA estimation using PIM with gap $D$ between two antennas equals to DoA estimation exploiting a ULA with $m$ antennas placed half wavelength apart, in which $m$ obeys (\ref{m/d}). This could lead to a huge reduction in complexity of the antenna array required for high precision DoA estimation -that reduces the required number of antennas from $m$ to 2-; if so, why is it not a common DoA estimation technique now? it is because DoA estimation using PIM is ambiguous and there are a number of different DoAs that can be inferred from a specific PIM \cite{vinci2011novel}; As $D$ increases CRLB decreases, however, ambiguity escalates. Moreover, DoA  estimation using PIM is not capable of detecting and discriminating between multiple concurrent DoAs. In section \ref{PS}, we propose a solution to estimate DoA using PIMs observed for multiple frequencies, instead of only measuring PIM for only a single frequency. We will see that this approach not only leads to PIM disambiguation, but also provides us with DoA estimation of all signal propagation paths between the source and the device. 

\subsection{Relationship Between DoA Estimation Precision, Beam-width and Resolution}
\label{DoABW}
In this section, we explain why DoA estimation precision and antenna beam-width are not necessarily
coupled, which further proves that spatial division (SD) and IA can be considered and performed as two completely independent tasks. Referring to (\ref{CRMPIM}) and (\ref{CRBULZ}), CRLB of angle estimation precision is directly related to SNR, as SNR increases precision improves; in other words, we can obtain any arbitrary precision if SNR is high enough regardless of $m$ or $D$. Although SNR can be improved by increasing the number of antennas, in a ULA, it can also be improved by integration, which is the time interval we can coherently receive and integrate a signal. Equivalently, angle precision can be improved only by integration, which come at a time cost, regardless of $m$ or $D$. 

\begin{figure}
    \centering
    \includegraphics[width=4.2in,height=3in]{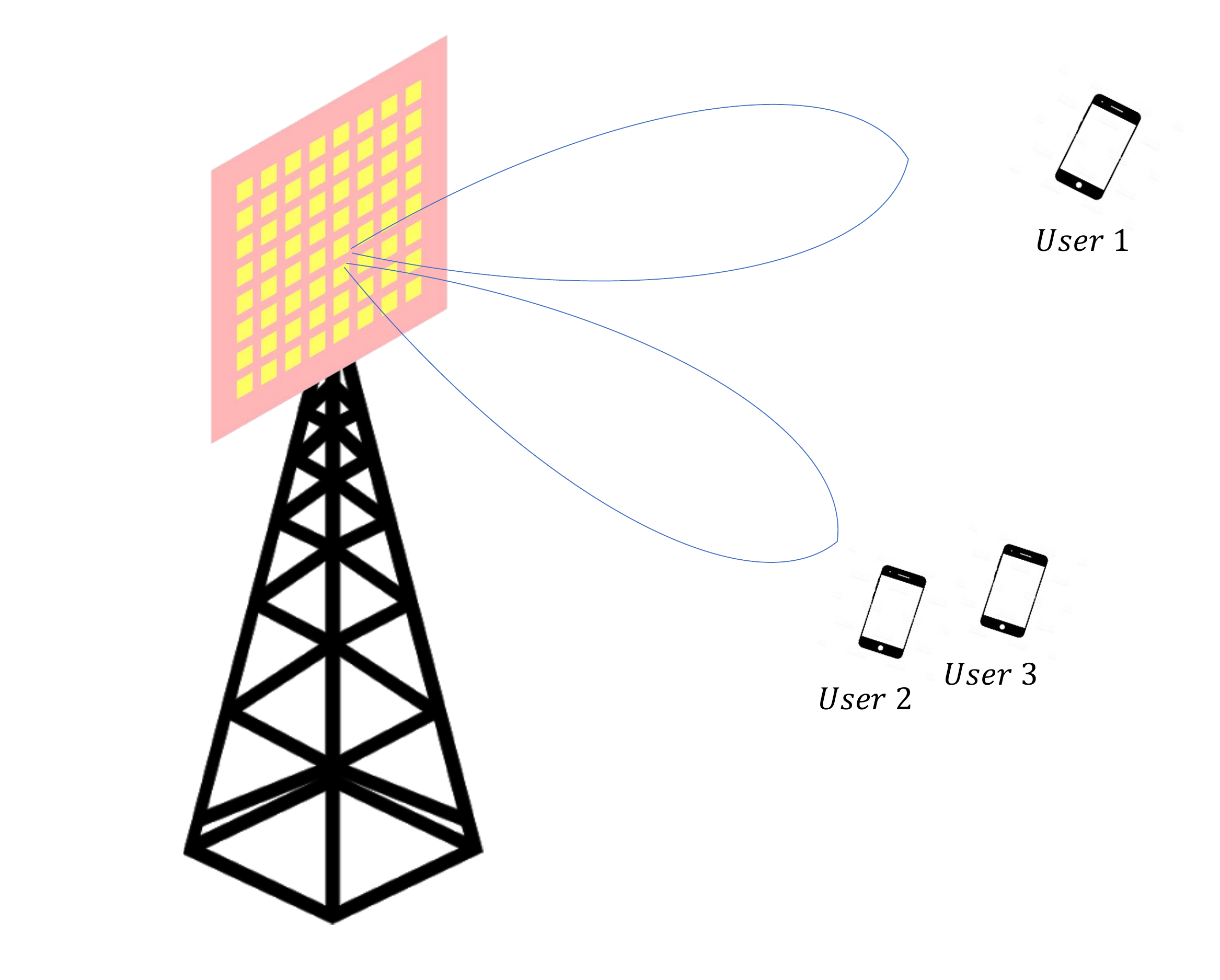}
    \caption{Visualisation of spatial division concept. The antenna is able to discriminate between user 1 and user 2,3 because the angular distance between them are more than beam-width. While, it is not able to discriminate between user 2 and user 3, since their angular distance is less than the antenna beam-width.}
    \label{SDMA}
\end{figure}

Now let's take a look at angle resolution concept. Angle resolution help us to measure the capability of a technique to discriminate between multiple incoming signals from different DoAs. We define angle resolution as the minimum angular distance between two incoming signals  that can be discriminated by a technique. Angle resolution is directly coupled with SD capability of a technique. A ULA can discriminate between two DoA if their angular distance is more than it's beam-width. Similarly, in the transmit mode, if the angular distance between two users is more than the beam-width and antenna sends signal to one of them, it causes much less interference for the second device compared to the situation where their angular distance are less than the beam-width (Figure \ref{SDMA}). In a ULA, beam-width is merely determined by the number of array elements and  equals to $\frac{2}{m}$. Therefore, the SD capability of a ULA is solely governed by its number of array elements.     

For ultra-fast mm-wave communication, devices has to be equipped with a highly directional antenna that enables SDMA. On the other hand, for initial access (IA), a good angle estimation is required. As we discussed earlier, an angle estimation with a desired precision can be obtained when SNR is high enough at the receiver. SNR and resolution are not two mutually-coupled aspects of a DoA estimation technique. Especially in the case of array antennas, resolution is governed by number of array elements, while DoA estimation precision is governed by SNR at the receiver. Consequently, we can seperate SD from IA, and dedicate a high processing gain technique for IA and a highly directional antenna for SD.

\section{Wide-band DoA Estimation Using Standing-wave Spectrometry}
\label{PS}
\begin{figure}
    \centering
    \includegraphics[width=6in,height=1.8in]{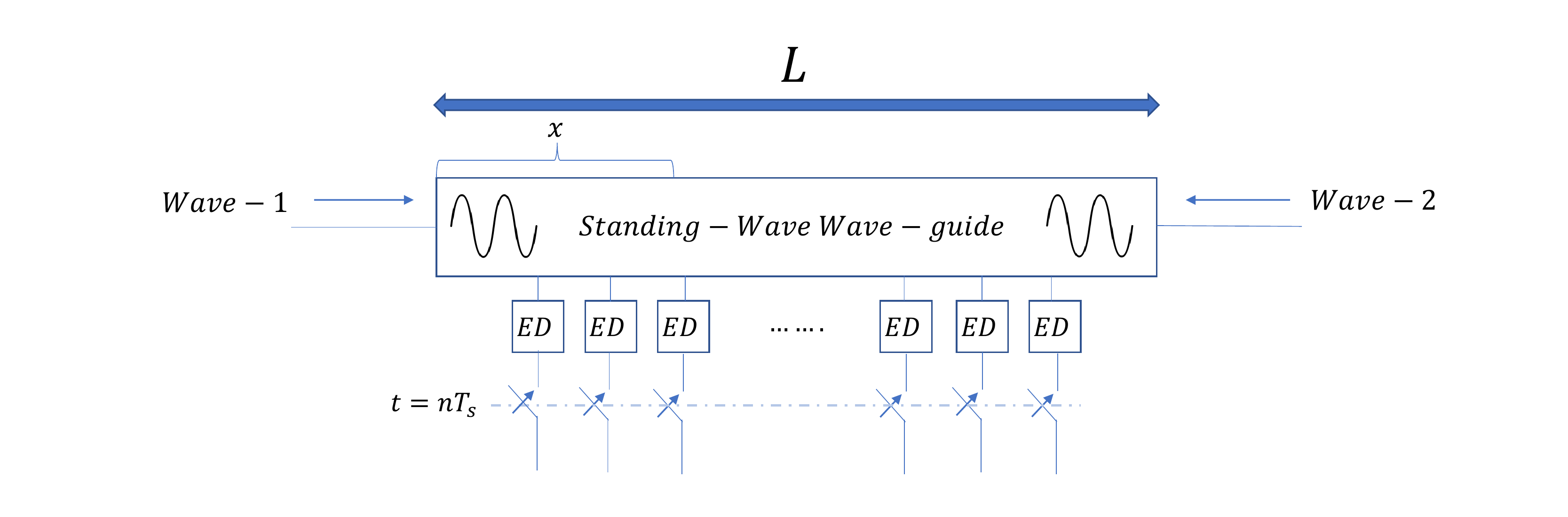}
    \caption{Standing-wave wave-guide. Two waves move in opposite directions interact to form a standing wave, the amplitude of the standing wave is sampled using a group of energy detectors (ED).}
    \label{SWR}
\end{figure}

In this section we inaugurate a new idea to estimate DoA of a signal using PDoAs. Here, we propose the source emits a signal with several gigahertz bandwidth in mm-wave, in such a way that the receiver can detect and discriminate between all (line-of-sight (LoS) and none-line-of-sight (NLoS)) paths between the source and the receiver using our proposed PS technique. Now suppose there exists $N_{NL}+1$ paths, 1 LoS path and $N_{NL}$ NLoS paths, between the source and the device. Given the source emits a monotone signal as of (\ref{sigmod}) with carrier frequency $f$ for the duration $T_p$, received signals at both antennas can be formulated as
\begin{equation}
\label{interefence}
     s^{(1)}_{R}(t) = \!\underbrace{a_0 e^{j 2\pi f t}}_{\text{LoS path}}\!+\!\underbrace{\sum_{k=1}^{N_{NL}}a_ke^{j 2\pi f (t-t_k)}}_{\text{NLoS paths}}+v_1(t)\;\;\; and\;\;\;
     s^{(2)}_{R}(t)=\!\underbrace{a_0 e^{j 2\pi f (t-\Delta{t_0})}}_{\text{LoS path}}\!+\!\underbrace{\sum_{k=1}^{N_{NL}}a_ke^{j 2\pi f (t-t_k-\Delta{t_k}))}}_{\text{NLoS paths}}+v_2(t)\,.
\end{equation}
where $t_k$ is the delay of signal arrival through NLoS path $k$ to the PIA w.r.t LoS path, and $\Delta t_k$ and $a_k$ is TDoA of signal between two antennas and amplitude of received signal through path $k$, $k = 0,\dots,N_{NL}$ (path 0 is the LOS path), respectively. Then, we guide the two received signals into a standing-wave wave-guide (SWWG) via two opposite directions (Figure \ref{SWR}). Referring to \cite{jovanov2010standing}, the first and the second paths of signal interact in the SWWG as 

\begin{align}
     &s^{(1)}_{R}(t) e^{j\beta(f) x} + s^{(2)}_{R} (t)e^{-j\beta(f) x}=\nonumber\\
     &e^{j 2\pi f t} \left(\left(a_0 e^{j\beta(f) x}+ a_0 e^{-j\beta(f) x}e^{-j 2\pi f (\Delta{t_0})} \right)+\left(\sum_{k=1}^{N_{NL}} a_k e^{-j 2\pi f t_k} \left(e^{j\beta(f) x}+e^{-j\beta(f) x} e^{-j 2\pi f (\Delta{t_k}))} \right)\right)\right)= \nonumber \\
     &e^{j 2\pi f t}\left(2a_0 e^{-j\pi f \Delta t_0}\cos{(\beta(f) x + \pi f \Delta t_0)}+\sum_{k=1}^{N_{NL}} 2 a_k e^{-j 2\pi f (t_k+\frac{\Delta t_k}{2})} \cos{(\beta(f) x + \pi f \Delta t_k)} \right) \,, \nonumber \\  
\end{align}

where $x$ is an arbitrary point along the SWWG, $L$ is the length of the wave-guide and $\beta(f)=\frac{2\pi} {\lambda_T}=2\pi \frac{f}{c_T}$, where $\beta(f)$, $\lambda_T$ and $c_T$ are phase constant, wavelength and phase velocity of electro-magnetive wave in the wave-guide, respectively \cite{steer2019microwave}. As Figure \ref{SWR} illustrates, using energy detectors along $x-axis$, we have
\begin{align}
    &E_{sw} (x,f) = \left| 2a_0 e^{-j\pi f \Delta t_0}\cos{(\beta(f) x + \pi f \Delta t_0)}+\sum_{k=1}^{N_{NL}}2a_k e^{-j 2\pi f (t_k+\frac{\Delta t_k}{2})} \cos{(\beta(f) x + \pi f \Delta t_k)} \right|^2 \nonumber \\
        &= \: 4 a^{2}_{0} cos^2 (\beta(f) x + \pi f \Delta t_0) +\sum_{k=1}^{N_{NL}} 4 a^{2}_{k} cos^2 (\beta(f) x + \pi f \Delta t_{k})\nonumber \\ 
    &+\sum_{k=1}^{N_{NL}} 8  a_0 a_k \cos{(\beta(f) x+ \pi f \Delta t_0)} \cos{ (\beta(f) x + \pi f \Delta t_k)} \cos {(2\pi f (t_k + \frac{\Delta t_k - \Delta t_0}{2}))} \nonumber\\ 
     &+\sum_{k=1}^{N_{NL}}\sum_{l=K+1}^{N_{NL}} 8  a_l a_k \cos{(\beta(f) x+ \pi f \Delta t_l)} \cos {(\beta(f) x + \pi f \Delta t_k)} \cos {(2\pi f (t_k-t_l + \frac{\Delta t_k - \Delta t_l}{2}))}\,.
     \label{Damaneh}
\end{align}

where $E_{sw} (x,f)$ is the output of the ED located at $x$. Interestingly, as \eqref{Damaneh} indicates, we could bypass down-conversion via mixing by using much simpler EDs. Now, suppose that input signal and its DoAs does not change during $T_p$, evidently, sampling rate after EDs can be as low as some $\frac{1}{T_p}$, if energy detectors provide energy integration of the wave for the whole duration. To simplify (\ref{Damaneh}), it is clear that $\Delta t_k \ll t_l$ and it is very probable that $\Delta t_k \ll t_k-t_l$;  $ k = 0,\dots,N, l = 1,\dots,N$  .
\footnote{The experimental results presented in \cite{rappaport2013millimeter} shows that delays of paths in two urban environment of New York and Austin is an order of several tens of nano seconds, on the other hand, the TDoA of signal between two antennas is a fraction of a nano second if the gap between antennas does not exceed $30 cm$.} Regarding \eqref{Damaneh}, the first and the second terms have \emph{cos(.)} components with parameters $\pi f \Delta t_0$ and $\pi f \Delta t_k$ , while the third and the forth terms have \emph{cos} components with parameters $\pi f t_k$ and $\pi f (t_k-t_l), k \neq l$, respectively. Given we measure \eqref{Damaneh} for multiple frequencies and $\Delta t_k \ll t_k-t_l$ for all $l,k$, applying Fourier transform over $E_{sw} (x,f)$ across $f$, the third and the forth terms of (\ref{Damaneh}) can be filtered out using a simple low-pass filter \footnote{In section \ref{DoARes} we will show that, this filter can be the same as the matched filter applied for DoA detection.}. The remaining terms after low-pass filtering are denoted by $ \hat{E}_{sw} (x,f)$

\begin{align}
   \hat{E}&_{sw} (x,f) = \: 4a_0^2 \cos^2 (\beta(f) x + \pi f \Delta t_0) +\sum_{k=1}^{N_{NL}} 4 a^2_k \cos^2 (\beta(f) x + \pi f \Delta t_k) \nonumber\\
   =2a_0^2& + 2 a^2 \cos  (2 \beta(f) x + 2\pi f \Delta t_0)+ \sum_{k=1}^{N_{NL}}2a^2_k + 2 a^2_k \cos (2\beta(f) x + 2\pi f \Delta t_k) \,.
   \label{phspecforf}
\end{align}

The number NLoS path from the source to the device are very few in mm-wave usually less than 3 path \cite{heath2016overview}, so $N_{NL} \le 3$. Here, if we estimate $\theta_k,a_k$ for $k = 0,\dots,N$, we can distinguish between all paths from the source to the device and determine signal received power from each path. In section \ref{DoARes}, \ref{longSW} and \ref{FreqCB}, we will discuss two different techniques that that can be used to detect DoAs based on sampling \eqref{phspecforf} in $f$-domain, and how the angular resolution that can be achived using PS.   

Consider that (\ref{phspecforf}) is derived by assuming a monotone signal is transmitted by the source. 
Now lets assume, signal is not monotone and has bandwidth $B$, thus signal can be expressed as

\begin{equation}
    s(t) =  \int_{f_c-\frac{B}{2}}^{f_c+\frac{B}{2}} a(f) e^{j2 \pi f t} df\,.
    \label{wide-spread}
\end{equation}

where $a_(f)$ is Fourier transform of $s(t)$. The received signals at the first and the second antennas can be expressed as

\begin{align}
\label{interefence1}
     &s^{(1)}_{R}(t)= \int_{f_c-\frac{B}{2}}^{f_c+\frac{B}{2}} a_0(f) e^{j 2\pi f t} df+\sum_{k=1}^{N_{NL}}\int_{f_c-\frac{B}{2}}^{f_c+\frac{B}{2}} a_k(f)e^{j 2\pi f (t-t_k)} df+v_1(t)\nonumber \\
     &s^{(2)}_{R}(t)=\int_{f_c-\frac{B}{2}}^{f_c+\frac{B}{2}}a_0(f) e^{j 2\pi f (t-\Delta{t_0})} df+\sum_{k=1}^{N_{NL}} \int_{f_c-\frac{B}{2}}^{f_c+\frac{B}{2}} a_k(f) e^{j 2\pi f (t-t_k-\Delta{t_k}))} df+v_2(t)\,.
\end{align}

Assuming a constant fading over $[f_c-\frac{B}{2},f_c+\frac{B}{2}]$, we can express $a_k(f)= \alpha_k a(f)$, where $\alpha_k$ denotes the attenuation of path $k$. Consequently, the two signals inside the SWWG can be formulated as \cite{jovanov2010standing}

\begin{align}
\label{interefence1}
     &s^{(1)}_{R}(t,x)= \int_{f_c-\frac{B}{2}}^{f_c+\frac{B}{2}} a_0(f) e^{j 2\pi f t} e^{j \beta(f) x} df+\sum_{k=1}^{N_{NL}}\int_{f_c-\frac{B}{2}}^{f_c+\frac{B}{2}} a_k(f)e^{j 2\pi f (t-t_k)} e^{j \beta(f) x}df+v_1(t)\nonumber\\
     &s^{(2)}_{R}(t,x)=\int_{f_c-\frac{B}{2}}^{f_c+\frac{B}{2}}a_0(f) e^{j 2\pi f (t-\Delta{t_0})} e^{-j \beta(f) x} df+\sum_{k=1}^{N_{NL}} \int_{f_c-\frac{B}{2}}^{f_c+\frac{B}{2}} a_k(f) e^{j 2\pi f (t-t_k-\Delta{t_k}))} e^{-j \beta(f) x} df+v_2(t)\,.
\end{align}
 Finally, the interaction between the two signals ($S_{int}$) in the SWWG can be formulated as 
 \small
 \begin{align}
     &S_{int}(t,x)=s^{(1)}_{R}(t,x)+s^{(2)}_{R}(t,x)=\nonumber\\
     &\int_{f_c-\frac{B}{2}}^{f_c+\frac{B}{2}}  \left(\left(a_0(f) e^{j\beta(f) x}+ a_0(f) e^{-j\beta(f) x}e^{-j 2\pi f (\Delta{t_0})} \right)+\left(\sum_{k=1}^{N_{NL}} a_k(f) e^{-j 2\pi f t_k} \left(e^{j\beta(f) x}+e^{-j\beta(f) x} e^{-j 2\pi f (\Delta{t_k}))} \right)\right)\right) e^{j 2\pi f t} df= \nonumber \\
     &\int_{f_c-\frac{B}{2}}^{f_c+\frac{B}{2}} \left(2a_0(f) e^{-j\pi f \Delta t_0}\cos{(\beta(f) x + \pi f \Delta t_0)}+\sum_{k=1}^{N_{NL}} 2 a_k(f) e^{-j 2\pi f (t_k+\frac{\Delta t_k}{2})} \cos{(\beta(f) x + \pi f \Delta t_k)} \right)e^{j 2\pi f t} df \,.   
\end{align}
\normalsize
Therefore, the power spectral density of $S_{int}$ turns out to be \cite{oppenheim2015signals}
 \small
\begin{align}
&\mathscr{{E}}_{sw}(x,f)=\lim_{T \to +\infty}\mathscr{F} \left\{ \frac{1}{T}\int_{0}^{T}|S_{int}(t,x)|^2dt\right\}\nonumber\\
&=\left|2a_0(f) e^{-j\pi f \Delta t_0}\cos{(\beta(f) x + \pi f \Delta t_0)}+\sum_{k=1}^{N_{NL}} 2 a_k(f) e^{-j 2\pi f (t_k+\frac{\Delta t_k}{2})} \cos{(\beta(f) x + \pi f \Delta t_k)} \right|^2 ; \: f \in [f_c-\frac{B}{2},f_c+\frac{B}{2}]\,.
\label{PSD}
\end{align}
\normalsize



As \eqref{PSD} shows $\mathscr{{E}}_{sw}$ exactly equals to \eqref{Damaneh} for $f \in [f_c-\frac{B}{2},f_c+\frac{B}{2}]$. Therefore, similar to the procedure of DoA estimation of a monotone signal, we can estimate all incoming signal DoAs and their power for a non-monotone signal using \eqref{phspecforf}. Now, lets again consider \eqref{phspecforf}, we can express $\hat{E}_{sw} (x,f)$ as summation of two terms
\begin{align}
   &\hat{E}_{sw} (x,f) = \sum_{k=0}^{N_{NL}} 2a_k^2  +\sum_{k=0}^{N_{NL}} 2 a^2_k \cos (2\beta(f) x + 2\pi f \Delta t_k) \nonumber\\
   &=\sum_{k=0}^{N_{NL}} 2a_k^2+\sum_{k=0}^{N_{NL}} 2 a^2_k \cos (\frac{4\pi}{c_T} fx + 2\pi f \Delta t_k)  \,.
   \label{phspecforfsimp}
\end{align}
Interestingly, factors $a_0,\dots,a_{N_{NL}}$ and phases $2\pi f \Delta t_0,\dots,2\pi f \Delta t_{{NL}}$, can simply be estimated by applying Fourier transform over $\hat{E}_{sw} (x,f)$ across $x$.  One useful example of signal as of (\ref{wide-spread}) is multiple single tones (e.g. 30 monotones) around the center frequency; in section \ref{FreqCB}, we show that this signal not only provides enough information to estimate all DoAs, but also enables integration to achieve very high SNRs, which results in very high precision DoA estimation.

\subsection{DoA Detection and Resolution}
\label{DoARes}
As we proved in the previous section, the interaction between two waves received at each antennas forms a standing-wave and its amplitude can be measured as of (\ref{phspecforfsimp}) for each frequency $f$, measuring amplitude of the standing-wave, employing a group of EDs. Consider that \eqref{phspecforfsimp} consists of $2 a^2_k \cos (2\beta(f) x + 2\pi f \Delta t_k)$ terms. Thus, estimating $\Delta t_k$ and $\alpha_k$ for $k = 0,\dots,N_{NL}$ is equivalent to harmonic decomposition of \eqref{phspecforfsimp} in $f$-domain. There are several techniques has been introduced for harmonic decomposition, such as Fourier transform, multiple signal classification (MUSIC) \cite{schmidt1986multiple}, Pisarenco harmonic decomposition \cite{pisarenko1973retrieval}, to name a few. Here for simplicity we only use a matched filter for DoA estimation. Given far-field assumption we have

\begin{equation}
    \Delta t_k=\frac{D\cos{\theta_k}}{c}\,.
\end{equation}

where $\theta_k$ is DoA of path $k$. Therefore, DoAs can be estimated applying the following matched filter on (\ref{phspecforf})

\begin{equation}
    h(\theta,f,x)=e^{j2\pi f \Delta t_k}e^{j2\beta(f)x}=e^{j2\pi f \frac{Dcos(\theta)}{c}}e^{j2\beta(f)x}=e^{j2\pi f (\frac{Dcos(\theta)}{c}+\frac{4\pi x}{c_T})}\,.
    \label{matchedfilter}
\end{equation}
(\ref{matchedfilter}) shows that the matched filter is a single monotone in the $f$-domain. Moreover, as $\Delta t_k$ increases, the matched filter represents a higher frequency signal in $f$-domain. Therefore, convolving (\ref{matchedfilter}) with (\ref{Damaneh}), the third and the forth terms of \eqref{Damaneh} will be eliminated. To calculate the angular resolution of PS suppose two different paths with two different DoAs $\theta_1,\theta'_1$ arrive at PIA and we can completely discriminate between $\theta_1$ and $\theta_1'$ using matched filter in (\ref{matchedfilter}), then we have 
\begin{align}
    &\int_{f-\frac{B}{2}}^{f+\frac{B}{2}} e^{j2\pi f D \frac{cos(\theta_1)-cos(\theta'_1)}{c}} df = 0 \rightarrow BD\frac{|cos(\theta_1)-cos(\theta'_1)|}{c}=k , k \in \mathbb{N} \rightarrow \frac{BD}{c} |\theta_1-\theta'_1||sin(\theta_1)|\approx k \nonumber \\
    &\rightarrow |\theta_1-\theta'_1| \approx \frac{ck}{BD |sin(\theta)|}
    \,.
    \label{msdsds}
\end{align}
Therefore the minimum possible angular distance between $\theta_1$ and $\theta_1'$ that can be resolved using our proposed technique (referred as DoA estimation resolution) can be approximated as
\begin{align}
    Res(\theta) \approx \frac{c}{BD |sin(\theta)|}
    \,.
    \label{Res}
\end{align}
Consequently, DoA estimation resolution is determined merely by $BD$, which means as the gap between two antennas or the signal bandwidth increases the DoA resolution will increase. As we mentioned earlier, in this we mainly use marched filter for DoA detection for simplicity, however, since PDoAs are available in digital domain, future works may consider more complicated signal processing techniques for DoA estimation. Those techniques may result in much better angular resolution than match filtering.

\subsection{Frequency Resolution}
\label{longSW}

Considering $\beta=2\pi \frac{f}{c_T}$, (\ref{phspecforfsimp}) clarifies that angle and phase difference of PIMs for any arbitrary frequency inside $[f-\frac{B}{2},f+\frac{B}{2}]$ would be easily extracted by applying Fourier transform over $\mathscr{{E}}_{sw}(x,f)$  across $x$, if we could measure $\mathscr{{E}}_{sw}(x,f)$ for an infinite length. Unfortunately, in practice we can only measure $\mathscr{{E}}_{sw}(x,f)$ for a limited length and it enforces a strong limitation on the frequency resolution of the Fourier transform. To calculate of resolution of FFT over $\mathscr{{E}}_{sw}(x,f)$ across $x$, consider that if we have a signal for length $T$ (in time), the highest FFT resolution possible is $\frac{1}{T}$ \cite{oppenheim1999discrete}. Given SWWG length is $L$, referring to (\ref{phspecforfsimp}), the frequency resolution ($\delta(f)$) turns out to be
\begin{align}
    Res(f)=\delta(f) \rightarrow 2 \frac{\delta f}{c_T} L =1 \rightarrow
    \delta(f)=\frac{c_T}{2L} \,.
    \label{Length}
\end{align}
Given $c_t\approx c$, to reach a $1GHz$ frequency resolution  we need a $15cm$ wave-guide and to reach a $100MHz$ resolution we need a $1.5m$ wave-guide. Such a long wave-guide may not be practical specially exploiting PCB or MMIC implementation since it results in a huge attenuation of the signal along the long wave-guide. Thus we may either employing alternative fabrication technologies or the following technique to resolve this issue. 
\subsection{Frequency Swiping Interferometry (Frequecny Code-book)}
\label{FreqCB}
Instead of spectrometry via a long wave-guide, we can sample PDoAs for a group of frequencies in $[f_c-\frac{B}{2},f_c+\frac{B}{2}]$ using a short wave-guide. To this end, we divide the frequency band into $S_f$ frequency steps (also referred as frequency code-book), each step is represented by a monotone (pilot), and measure (\ref{phspecforf}) for each pilot. We also divide the whole PS duration into $S_f$ time slots and measure PDoA for each pilot at each time slot. Since we measure PDoA for a monotone in each time slot, our approach bypasses the need for a long SWWGs. Consider that, the number of pilots and the distance between them (in $f$-domain) should provide us enough information to detect all DoAs. Referring to (\ref{phspecforf}), we measure $e^{j 2\pi f \Delta t_{in}}$ for each pilot, where $t_{in}$ can potentially changes between $[-\frac{D}{c},\frac{D}{c}]$, therefore we should sample the phase difference with at least $\frac{c}{2D}$ rate (Nyquist rate) in the $f$-domain to capture all information regarding $\Delta t_{in}$, thence, the code-book should contain at least 
\begin{equation}
    \mathrm{min} \: S_f= \frac{B}{\frac{c}{2D}}=\frac{2BD}{c}\,,
\end{equation}
pilots (samples in $f$-domain). Consequently, we propose to establish a directional link between two devices, both devices should send pilots, so the other side can estimate DoAs of signal based on measuring PDoAs for all pilots. Using our proposed technique, there is no need for spatial search and all DoAs can be estimated via measuring PDoAs of pilots. Lets $f_0$ denotes the frequency of the first monotone and $\Delta f_0 = \frac{c}{2D}$ denotes the distance between pilots in $f-$domain. Thus, the vector of all measured phases for the frequency codebook ($\boldsymbol{\Delta \phi}$) can be expressed as 

\begin{align}
&\boldsymbol{\Delta \phi}=
  \begin{bmatrix}
      e^{j 2\pi f_0 \Delta t_{in}} & e^{j 2\pi (f_0+\Delta f_0) \Delta t_{in}} & \dots & e^{j 2\pi (f_0+(S_F-1)\Delta f_0) \Delta t_{in}} 
\end{bmatrix} 
\nonumber \\
&=e^{j 2\pi f_0 \Delta t_{in}}
\begin{bmatrix}
        1 & e^{j 2\pi (\Delta f_0) \Delta t_{in}} & \dots & e^{j 2\pi ((S_F-1)\Delta f_0) \Delta t_{in}}  
\end{bmatrix} \,,
\label{PSULASIM}
\end{align}
where $\Delta t_{in}$ is the TDoA of signal between two antennas. \eqref{PSULASIM} equals to the vector of phase differences measured by a ULA ($\boldsymbol{\Delta \phi_{u}}$) with $S_F$ elements multiplied by $e^{j 2\pi f_0 \Delta t_{in}}$  
\begin{align}
&\boldsymbol{\Delta \phi_{u}}=
  \begin{bmatrix}
      1 & e^{j 2\pi (\Delta f_0)\Delta t_{d}} & \dots & e^{j 2\pi ((S_F-1)\Delta f_0) \Delta t_{d}} 
\end{bmatrix} \,,
\label{ULAPSH}
\end{align}
where $\Delta t_{d}$  is the TDoA between of signal between two consecutive elements and $\Delta f_0$ is the working frequency of ULA. In fact, we reconstruct a ULA that works at frequency $\Delta f_0$ via PS that works at much higher frequency $f_0$ \footnote{More interestingly, $f_0$ and $\Delta f_0$ are independent. $f_{0}$ should be high enough to provide us with enough unused bandwidth required to emulate the ULA. Thus, PS is much more applicable in mm-Wave and Terahertz bands becuase large swaths of spectrum is available.}. As $B$ increases the number of pilots (equivalent to ULA elements) can increase and as $D$ increases $\Delta f_0$  decreases and again we can increase the number of pilots which results in better angular resolution. In a ULA, usually PDoAs of \eqref{ULAPSH} are compensated by phase shifters at each elements for different values of possible $\Delta t_{d}$s to find the best match with $\boldsymbol{\Delta \phi_{u}}$ and detect the DoA (i.e. the spatial search). In our technique, since we measure PDoAs using PS techniques we can find the incoming DoA by digital signal processing. In section \ref{sim} we will show that output of matched filter of \eqref{matchedfilter} applied on \eqref{PSULASIM} is very similar to output of phase shifters applied on \eqref{ULAPSH} (conventional beamforming). \footnote{Throughout this work, we only consider a simple matched filter on \eqref{PSULASIM} to detect DoAs. However, PS provides \eqref{PSULASIM} in the digital domain, thus, much more complex signal processing techniques can be applied. Future works may consider various frequency sampling and corresponding array signal processing techniques to improve PS performance.}  
\subsection{SNR Analysis}
\label{SNR}

\begin{figure}
    \centering
    \includegraphics[width=6in,height=3.5in]{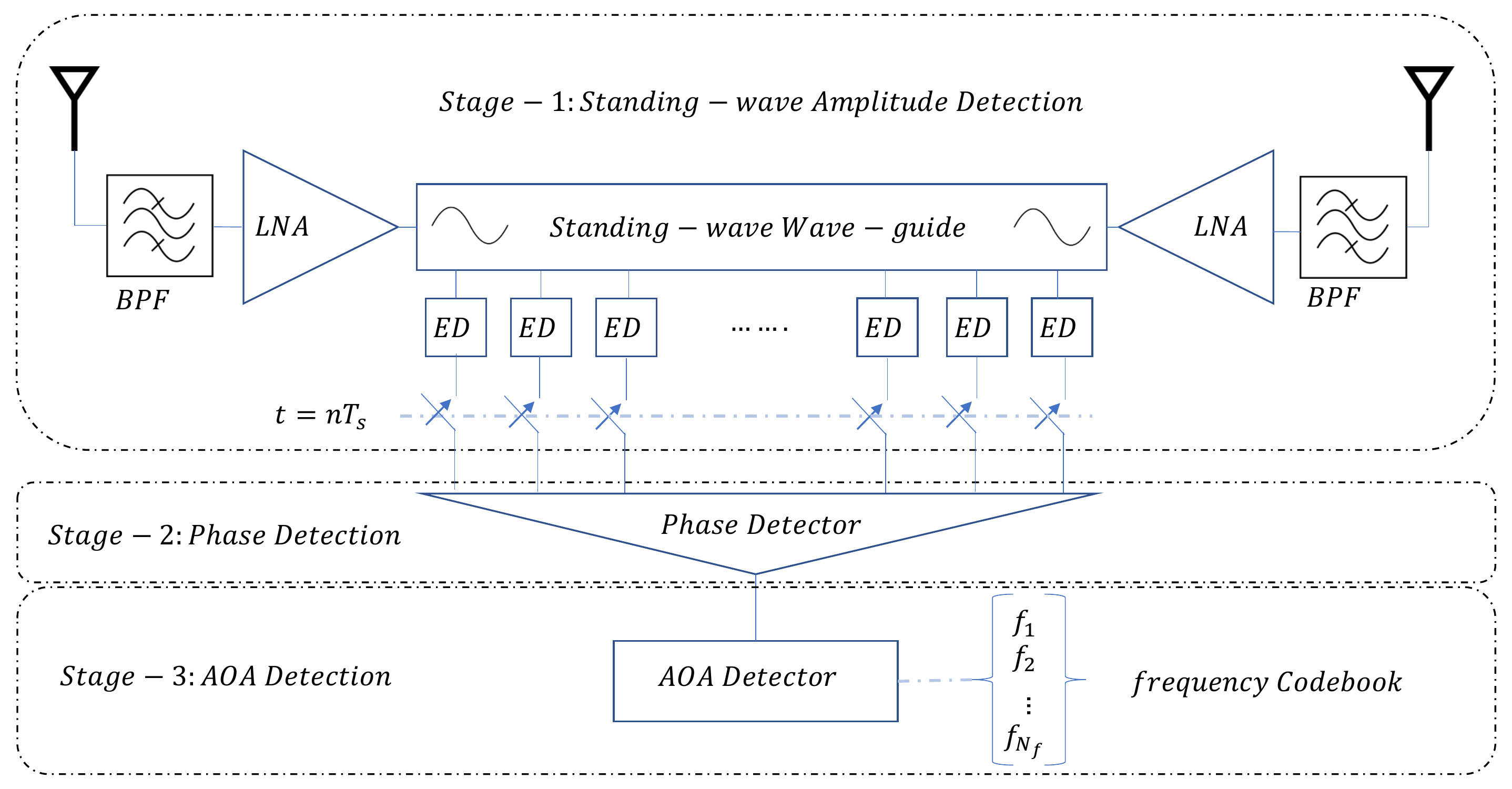}
    \caption{Block diagram of our proposed DoA estimation technique. Input SNR is improved through stages 1 and 3. In stage-1, a monotone signal is received at two antennas and passes through a band-pass-filter (BPF) via each path. After amplification via a low-noise-amplifier (LNA) in each path, both signals enter a wave-guide to form a standing-wave. Amplitude of the standing-wave is measured by a group of energy detector (ED) sensors, which inherently are low-pass filters and therefore, improves the SNR. Then the amplitude is sampled and can be integrated during each monotone time-step ($T_p$). After sampling, signal passes through a phase detector. Finally, PDoAs measured for all frequencies of the code-book are used to estimate DoAs using a matched filter which improves SNR for the second time.}
    \label{BD}
\end{figure}

Long SWWGs is subject to suffering from a huge loss, specially in mmwave. Since SNR is an absolutely critical factor when we deal with millimeter waves, it is more practical not to attenuate the input signal in the receiver by employing long SWWGs. In this section we analyse SNR of the technique that employs a frequency code-book instead of a long SWWG. The block diagram of the receiver using the frequency code-book technique is depicted in Figure \ref{BD}. As the figure illustrates, input signals pass through 3 stages until DoAs of signal are detected. Each stage may improves SNR. To measure how much the proposed receiver improves SNR we use the processing gain ($G_p$) metric \cite{rouphael2009rf}. Procesing gain is defined as ratio of the SNR of a processed signal to the SNR of the input signal. $G_p$ of the whole receiver can be expressed as

\begin{equation}
    G_p(total)=G_p(stage-1)G_p(stage-2)G_p(stage-3)\,.    
\end{equation}

Now lets calculate the $G_p$ for each stage. We ignore losses caused by hard-wares in our calculation. Consider a very basic formula that governs $G_p$ of any arbitrary process \cite{dixon1994spread}
 
\begin{equation}
G_p=\frac{B_{rf}}{B_{info}}=B_{rf}T_{int}\,,
\label{GP}
\end{equation}

where $B_{rf}$ in input bandwidth, and $B_{info}$ is the information bandwidth and $T_{int}$ is the integration time. This formula states that you can improve SNR of the input signal by integration as long as noise of samples are independent, otherwise integration will amplify the noise the same as signal and SNR won't improve. To make it more clear, suppose that input signal bandwidth is $1 MHz$, and assume that it is sampled by $ 1MHz$ sampling rate. Then we integrate the signal coherently for $1 ms$, in other words, we integrate $1000$ samples of the signal coherently. Consequently, $G_p=1000=\frac{1Mhz}{1Khz}=1Mhz*1ms$. If we sample the signal with a higher sampling rate, we will have more samples for integration, however, noise of samples are correlated and the integration won't result in higher SNRs.  
In view of (\ref{GP}), lets calculate $G_p$ for the first stage. Given each monotone of the code-book is received for $T_p$, assuming bandwidth of $B_{rf}$ for the BPF, $G_p$ of the first stage can be formulated as, 
\begin{equation}
    G_p(stage-1)=B_{rf}T_p\,.
\end{equation}
Consider that the only information that each ED measures is the amplitude of the standing wave, which is constant during $T_p$, Therefore, the amplitude can be estimated by integrating the input signal for $T_p$.
To calculate $G_p$ for the next stage, consider that the wave-guide length is $L$ which is in order a wavelength, as we sample the standing wave through the wave-guide, it is equivalent to sample the standing wave in time with a rate more than $f_c$, since $B_{rf}$ is much less than $f_c$, noise of these samples are not independent and integration at the second stage won't result in  any SNR improvement. 
At the last stage we measure PDoAs for the frequency code-book in different time slots, therefore noise of phase difference measurements at each time slot is independent of all other time slots -even if frequencies of pilots at two different
time slots are the same-; therefore, PDoAs can be integrated over all the code-book's pilots and the processing gain of stage-3 can be expressed as     
\begin{equation}
   G_p(stage-3)=S_f\,.
\end{equation}%
Finally, the total $G_p$ (processing gain) of all stages is 

\begin{equation}
    G_p(total)=B_{rf}T_p S_f\,.
    \label{TPG}
\end{equation}

$T_p S_f$ equals total time spent on receiving pilots by the receiver, in other words, using the proposed technique, we can make use of the whole duration of DoA estimation procedure to improve input SNR and consequently, improve DoA estimation precision. As we discussed earlier, directional techniques spend substantial amount of time for spatial search to find the other side of the link, moreover, both sides can not search for each other at the same time which further increases the spatial search duration. Contrarily, employing our proposed technique, both sides are able to search for the other side at the same time and can take advantage of the whole search duration to improve DoA estimation precision.


\subsection{Uplink and Downlink DoA Estimation}
\label{scale}
\begin{figure}
    \centering
    \includegraphics[width=7in,height=3.5in]{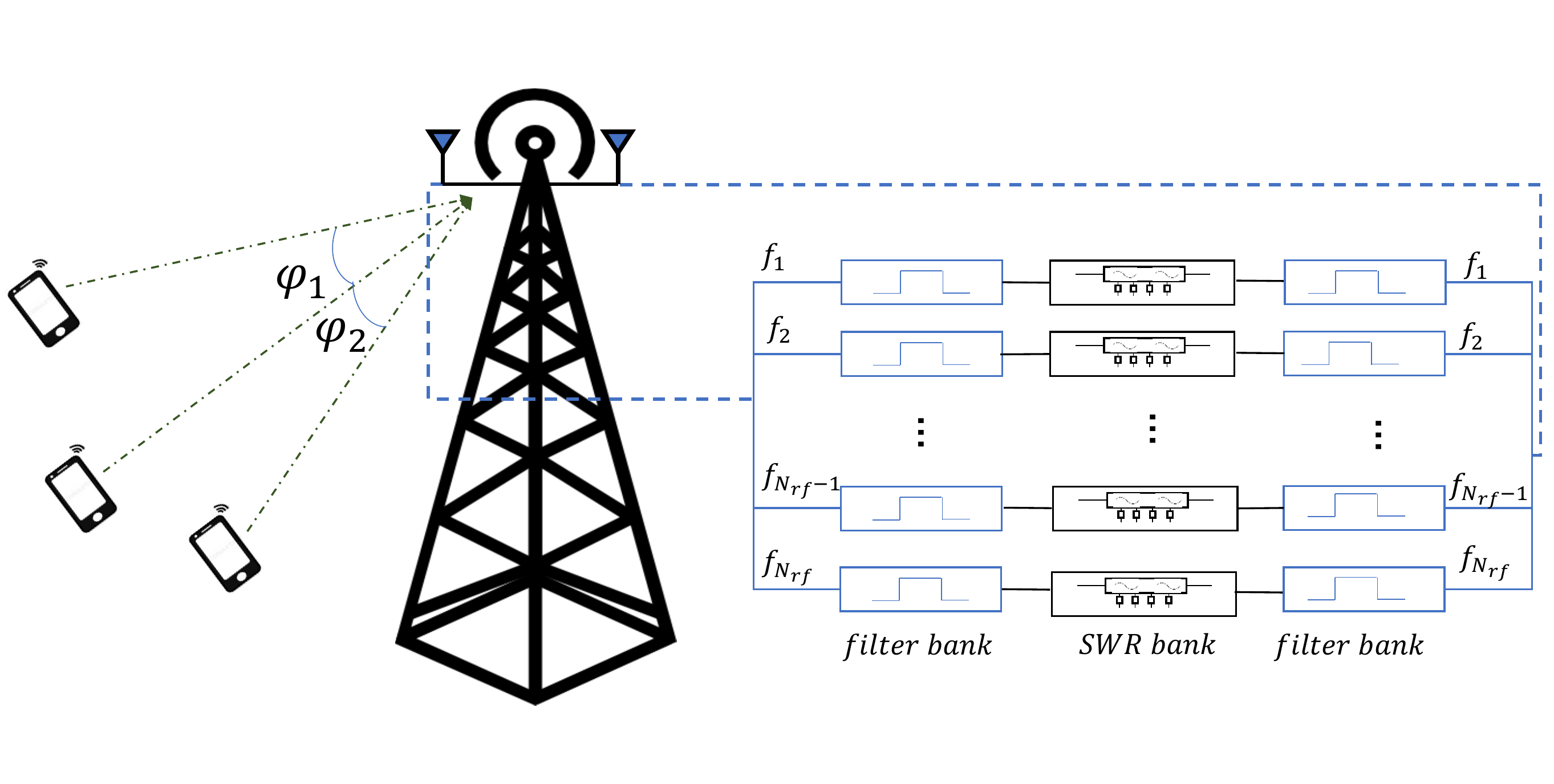}
    \caption{In uplink scenario, to be capable of discriminating between DoAs of multiple devices, the BS requires to be equipped with two filter-banks at both lines of it's PIA, and a separate SWR for each frequency sub-band.}
    \label{Uplink}
\end{figure}

In this section we are going to answer the following question: "How does PS perform in the presence of multiple users? How many devices can find their relative angles simultaneously using PS?" to answer these questions assume the following scenario: There is a base-station (BS) and $N_d$ devices around it in an environment, all devices require to estimate signal DoAs from the base-station (downlink), and the base-station requires to know DoAs of signals from devices (uplink). In downlink scenario, it is only required that BS sends one common code-book and all devices can find DoA of BS by measureing PDoAs of pilots of the common code-book. However, the uplink scenario is more complicated. If all  devices send the same code-book it is impossible for the BS to distinguish between DoAs. Therefore, devices' code-books have to be orthogonal either in time or frequency. If the BS can split the code-book band ($B$) to $N_{rf}$ sub-bands and uses an exclusive SWR for each sub-band, it can estimate DoA from $N_{rf}$ devices simultaneously (Figure \ref{Uplink}), since, $N_{rf}$ different frequency code-books can be processed simultaneously at the BS. Considering, the BS can be equipped by antennas with much larger $D$ and more complicated receivers than devices, the BS can estimate DoA from multiple devices simultaneously. 

\subsection{Ultra-fast DoA Estimation}
\label{UFast}
As we discussed in section \ref{FreqCB}, we suggest measuring PDoAs for multiple frequencies over multiple time slots to avoid using a large SWWG. In that architecture, we assumed we can only use a single SWR. Therefore, we have to measure PDoA of different pilots at different time slots. Nevertheless, instead of using a single SWR, it is possible to use a cascade of multiple SWRs, discriminating between multiple pilots using a filter bank, and find the PDoA for each monotone exploiting a specific SWR (the architecture is presented in Figure \ref{Uplink}). Using such an architecture, we can estimate all incoming DoAs in a single time slot, without any negative impact on the processing gain and the DoA estimation precision. Such an ultra-fast DoA estimation has not been previously possible using directional antennas, since those techniques are bound to spatial search. Ultra-fast DoA estimation using PS requires more complex hardwares in comparison to the  technique introduced in section \ref{FreqCB}, which may make it overpriced or oversized to be implemented on commercial mobile phones. However, it may be very promising for applications such as radar, mm-wave network backhaul, UAV and satellite communications, where more complex and bulky hard-wares can be implemented on devices. 

\section{Simulation Results}
\label{sim}

In this section, the perfromance of the proposed DoA estimation technique for different parameters is studied.
 \subsection{Simulation Setup and Results}
 
\begin{figure}
    \centering
    \includegraphics[width=5in,height=3.5in]{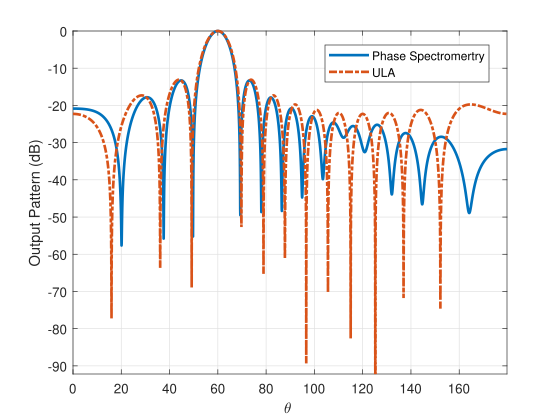}
    \caption{The matched filter of (\ref{matchedfilter}) is applied to phase differences measured for 40 pilots of a frequency code-book that changes between $[55,65] GHz$ and the output is plotted for $\theta$ between $[0,180]^o$ and is compared with a beam pattern of a ULA with 13 elements \cite{er1990linear}. $DoA_{in}=60^o$, $f_c=60Ghz$, $SNR_{in}=20dB, \frac{BD}{c}=6.67$. $D=20cm$ }
    \label{simfig1}
\end{figure}

 In the first simulation, we examine a basic scenario where a signal arrive at PIA through only one path, therefore, there is only one DoA to be estimated. We set $f_c=60 GHz$, $B=10GHz$, the steps of the codebook is 40 and pilots are selected equally spaced from $55 GHz$ to $65 GHz$ and $T_p=1 \mu s$, $c=3 * 10^8 \frac{m}{s}$, $D = 20 cm$, $L=2.5 mm$ and the number of EDs along the SWWG is set to 30. The received $SNR$ in each antenna is set to $20dB$ and the DoA of the signal is set to $60^o$. Figure \ref{simfig1}  shows the result of applying the matched filter of (\ref{matchedfilter}) for differnt $\theta$. As Figure \ref{simfig1} illustrates the output shows a distinctive peak at $60^o$. Moreover, Figure \ref{simfig1} illustrates that PS output pattern is similar to beam-pattern of a ULA with 13 elements. This result may seem contradictory to (\ref{m/d}), which indicates that FIM of angle estimation using PIMs equals to a FIM of a ULA with $m$ elements, in which $m$ obeys (\ref{m/d}), that results in $m=33$ applying the mentioned parameters. Keep in mind that, (\ref{CRMPIM}) shows CRLB of angle estimation using PIMs if and only if signal from one source is received at PIA, on the other hand, Figure \ref{simfig1} shows how PS can discriminate between two or more signals if they are originated from different DoAs. As (\ref{CRMPIM}) indicates, this bound is only a function of $D$ and SNR, while (\ref{Res}) shows that DoA estimation resolution is a function of $BD$, which means our technique can discriminate between two incoming DoAs if and only if $B$ is wide enough.                
 
\begin{figure}
    \centering
    \includegraphics[width=7in,height=5in]{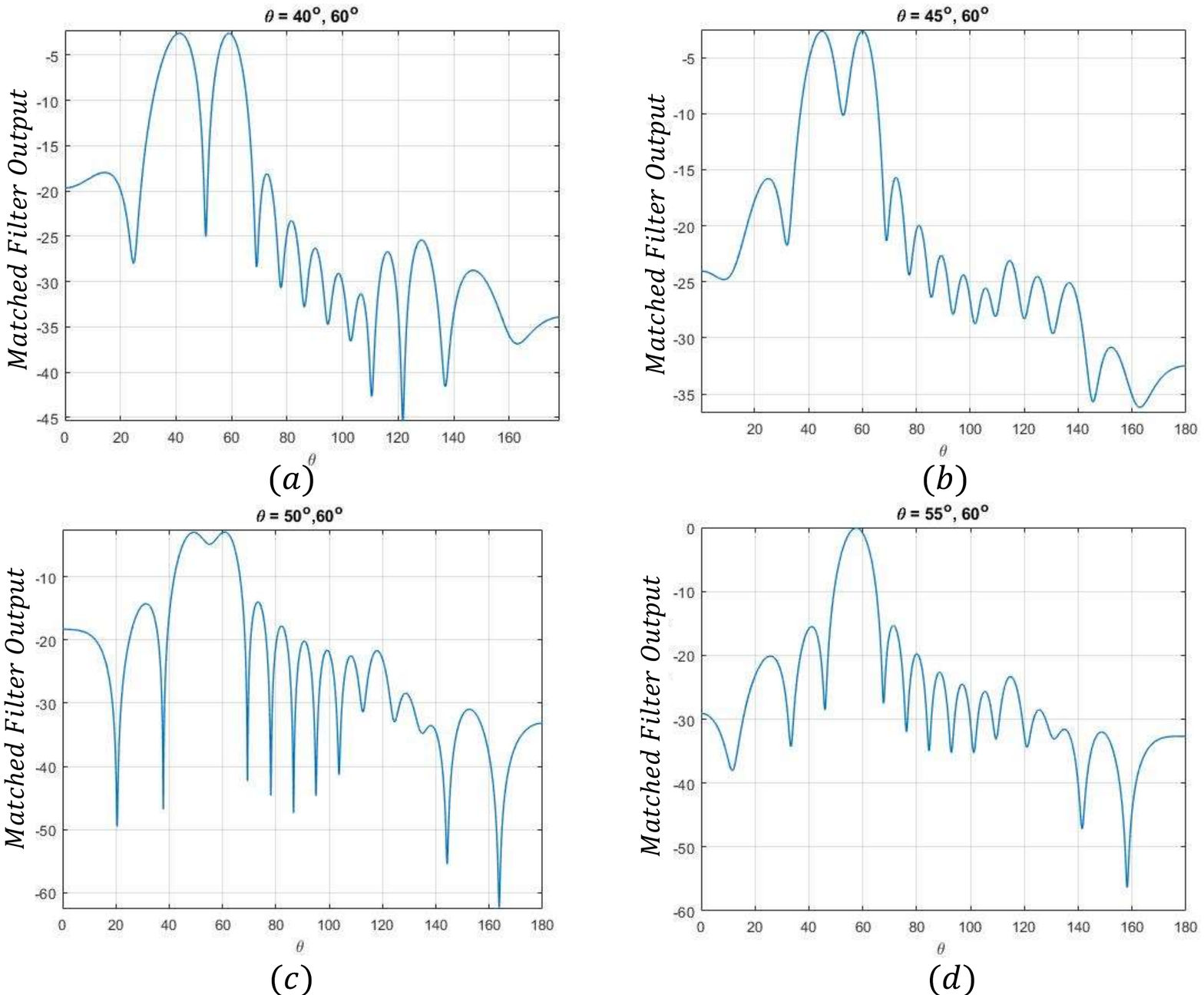}
    \caption{To analyse DoA estimation resoltion, the matched filter outputs are depicted for 4 different pairs of incoming DoAs : (a) $40^o,60^o$, (b) $45^o,60^o$, (c) $50^o,60^o$, (d) $55^o,60^o$. $SNR_{in}=20dB, \frac{BD}{c}=6.67$. }
    \label{simfig2}
\end{figure}

\begin{figure}
    \centering
    \includegraphics[width=4.5in,height=3in]{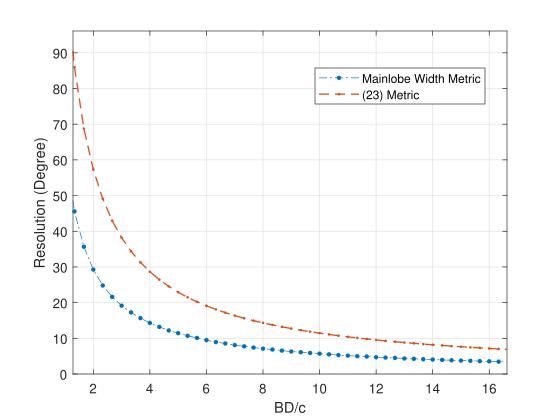}
    \caption{DoA estimation resolution based on the main-lobe width metric and the meric introduced by (\ref{Res}). $SNR_{in}=20dB$,    $DoA=60^o$ }
    \label{simfig4}
\end{figure}

 In the following simulation we are going to study DoA estimation resolution of the technique. In this simulation, parameters are the same as the first simulation, unless, we assume that the signal received at PIA from two different paths and two different DoAs, we investigate whether the proposed technique can distinguish between these two DoAs or not. Figure \ref{simfig2} shows the matched filter output for 4 different pairs of DoAs, the gap between 2 DoAs are $20^o,15^o,10^o,5^o$ respectively. As Figure \ref{simfig2} illustrates, when the gap between two DoAs is $20^o$, two lobs regarding each DoA are completely separated and distinguishable. When the gap resuces to $15^o$, two lobs start merging together, however, two peaks regarding two DoAs are again distinguishable. As the gap further reduces to $10^o$, two lobes merges more and two peaks are hardly distinguishable. And finally when the gap reduces to $5^o$, two lobes completely merge together and two peaks are not distinguishable. With respect to (\ref{Res}), the DoA resolution with $B=10Ghz$ and $D=20cm$ is approximated to be $17^o$. Since we calculate (\ref{Res}) assuming matched filters of two DoAs are perpendicular to each other, which means that two lobes are completely separated, thus simulations results are in compliance with (\ref{Res}). However, it seems that it is a strict metric for DoA resolution, to assume that two DoA are resolvable only if two lobes are completely separated. In practice, we may use $75\%$ or $50\%$ of (\ref{Res}) as a more realistic metric of the resolution. In Figure \ref{simfig4}, we illustrate matched filter main-lobe width and (\ref{Res}) versus the parameter $\frac{BD}{c}$, given $DoA=60^o$. Main-lobe width is defined as the gap between the minimum and the maximum $\theta$ in which the matched filter output is closer than 3db to its peak. As Figure \ref{simfig4} expresses, main-lobe width for $\frac{BD}{c}=6.67$ is $8.5^o$ which is half of the figure calculated by (\ref{Res}), moreover, this proportion between main-lobe width and (\ref{Res}) almost holds for every $\frac{BD}{c}$. Therefore we can use half of (\ref{Res}) as the DoA estimation resolution if we consider the more practical main-lobe width metric.                         
 
\begin{figure}
    \centering
    \includegraphics[width=4.5in,height=3in]{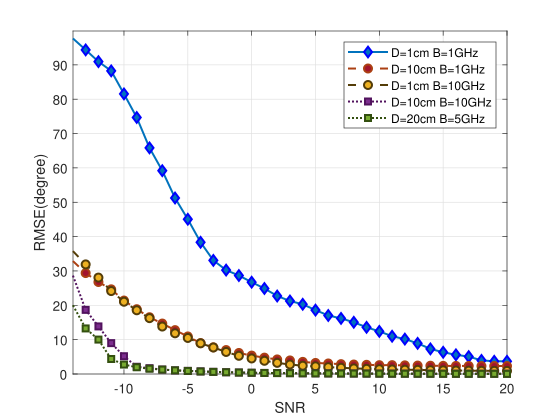}
    \caption{RSME of DoA estimation versus input SNR for differnt valus of $B$ and $D$. $DoA=60^o$, $S_f=40$,$B_{rf}T_p=100$.} 
    \label{simfig3}
\end{figure}

\begin{figure}
     \centering
     \begin{subfigure}[b]{0.4\textwidth}
         \centering
             \includegraphics[width=1.2\textwidth]{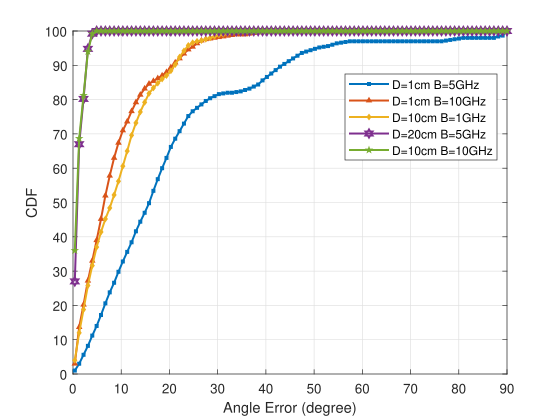}
         \caption{SNR=-10dB}
         \label{CDF1}
     \end{subfigure}
     \hfill
     \begin{subfigure}[b]{0.4\textwidth}
         \centering
             \includegraphics[width=1.2\textwidth]{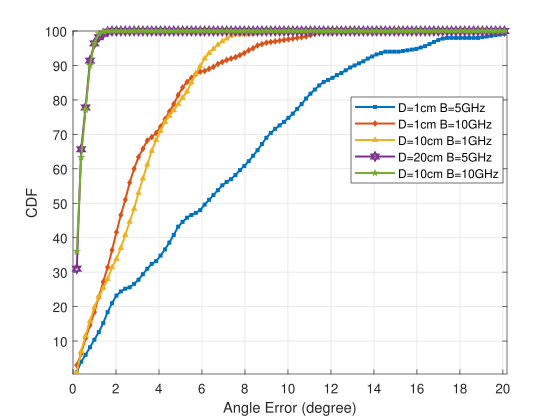}
         \caption{SNR=-5dB}
         \label{CDF2}
     \end{subfigure}
     \hfill
     \begin{subfigure}[b]{0.4\textwidth}
         \centering
         \includegraphics[width=1.2\textwidth]{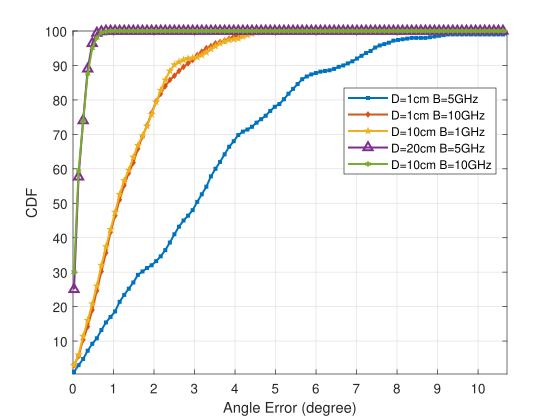}
         \caption{SNR=0dB}
         \label{CDF3}
     \end{subfigure}
     \hfill
     \begin{subfigure}[b]{0.4\textwidth}
         \centering
         \includegraphics[width=1.2\textwidth]{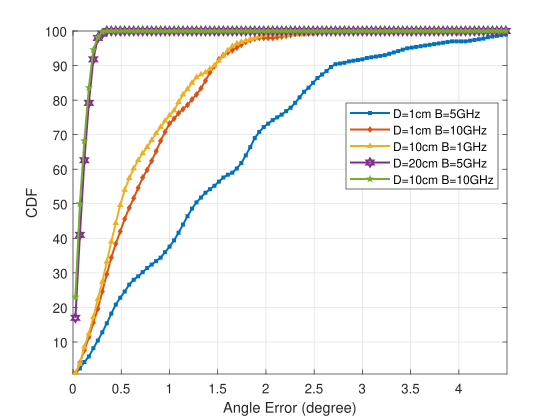}
         \caption{SNR=5dB}
         \label{CDF4}
     \end{subfigure}
        \caption{CDF of angle estimation error for differnt values of $D$, $B$ and input SNR. $DoA=60^o$, $S_f=40$,$B_{rf}T_p=100$.}
        \label{CDF}
\end{figure}

In the next simulation we analyse the effect of input SNR on DoA estimation error for differnt values of $B$ and $D$. In this simulation, input SNR changes in the interval $[-15,20]dB$. To analyse the error we calculate the root-mean-square error (RMSE) for each input SNR, by repeating the simulation 1000 times and find the average of SE for each SNR. Figure \ref{simfig3} illustrates RSME of DoA estimation error. As Figure \ref{simfig3} illustrates angle estimation error depends on $BD$, as $BD$ and SNR increases, error declines.  Similarly, Figure \ref{CDF} shoes that error CDF of PIAs with equal $BD$ factors are roughly the same. This is consistent with our results on angle resolution. However, it may seems inconsistent with (\ref{CRMPIM}), which indicates that CRLB of DoA estimation decrease in proportion to $D$ not $BD$, this is because we employ matched filter of (\ref{matchedfilter}) to find the DoA. To improve the precision, future works may considering using the output of the matched filter only to disambiguate the phase to a valid TDoA and estimate DoA directly based on the TDoA. 

\begin{figure}
    \centering
    \includegraphics[width=4.5in,height=3in]{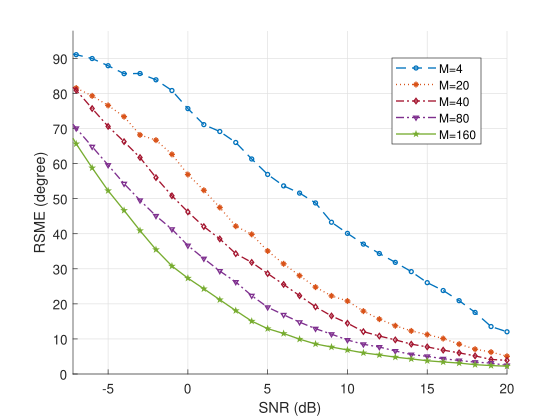}
    \caption{RSME of DoA estimation. Source only transmits four pilots at $[59.5,59.83,60.16,60.5]GHz$, each in a time slot with duration $T_p$, source repeats emitting these monotones for $M$ time slots. $DoA=60^o$, $T_p=100ns$, $B_{rf}=100MHz$, $D=1cm$.} 
    \label{tightcon}
\end{figure}

In the next simulation we consider a scenario in which, frequency steps, band-width, antenna gap and integration time is strongly limited. In this scenario, the source can only send 4 pilots at $[59.5,59.83,60.16,60.5]GHz$, $D=1cm$, $T_p=100ns$ and the whole number of available time slots is $M$. The source send those four frequencies in $M$ time slots respectively and repeats sending them until covers the whole $M$ slots. Consequently, the integration time is $MT_p$ -the maximum integration time in this simulation is $16 \mu s$-. We also assume there is only one incoming DoA at the PIA, since the PIA is not able to discriminate between two DoAs because of limited bandwidth and short antennas' gap. As Figure \ref{tightcon} shows, the proposed technique is able to estimate DoA with RMSE less than $10^o$ if input SNR is high enough, for $M=160$ input SNR should be above $7dB$ and for $M=20$ input SNR should be above 16dB. Therefore, as input SNR levels decreases we should increase integration time of our technique to provide us with acceptable DoA estimation precision.               

\begin{figure}
    \centering
    
    \includegraphics[width=4.5in,height=3in]{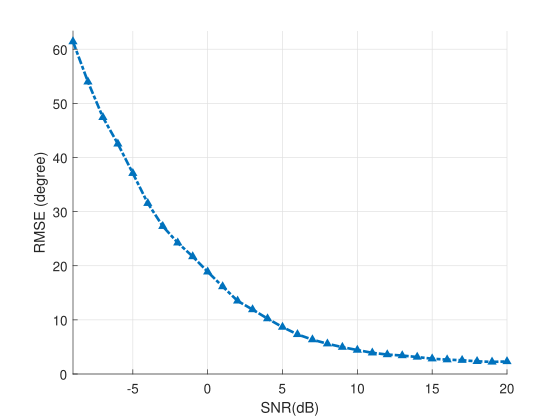}
    \caption{RSME of DoA estimation, signal receives at PIA via two paths, one LoS path ($90^o$) and one NLoS path ($30^o$) in the presence of coherent interference. Source only transmits four pilots at $[59.5,59.83,60.16,60.5]GHz$. $SIR=15dB$, $M=400$, $DoA=60^o$, $T_p=100ns$, $B_{rf}=100MHz$, $D=1cm$.} 
    \label{tightconSIR}
\end{figure}

In the next simulation, parameters are the same, unless there is a NLoS path ($30^o$) besides the LoS ($90^o$) path with a power 15 dB less than LoS path. This simulation is consistant with the experimental results of \cite{rappaport2013millimeter} on distribution of DoA paths between TX and RX in an urban environment in Brooklyn, New York. In This simulation integration time is set to $40 \mu s$. As Figure \ref{tightconSIR} shows existence of the second path does not have a considerable effect on RMSE of the proposed technique. Therefore, it seems that even a very simplified version of the proposed technique (narrow beam-width, short antenna gap) can be used in real world practical mm-wave DoA estimation applications.

In the next simulation, we investigate DoA estimation precision based on power of NLoS path. Given LoS path arrives at $90^o$ and NLoS path arrives at $30^o$ at the PIA, Figure \ref{tightconSIR} depicts RMSE of DoA estimation versus power ratio of LoS path to NLoS path. we set the integration time to be $4 \mu s$, since NLoS path can be considered as a coherent interference, thus SIR won't be improved by integration.  Figure \ref{tightconSIR} expresses that RMSE drops below $10^o$ when SIR is higher than $8dB$ and $5^o$ when SIR in higher than $12dB$. Referring to \cite{rappaport2013millimeter}, the  power of the strongest NLoS path expects to be more than 15dB weaker than the LoS path in a dense urban environment, therefore we expect that the proposed technique can estimate DoA of LoS path in an urban environment with error less than $3^{o}$ even when the available bandwidth is very limited (e.g. 1GHz) and the antenna gap is very short (1cm). Such a performance make PS a promising technique for beam initialization requirements of 5G networks, since the required band-width is easily accessible in mm-wave and the PIA size is very small that make it easily implementable on any device. 

\begin{figure}
    \centering
    \includegraphics[width=4.5in,height=3in]{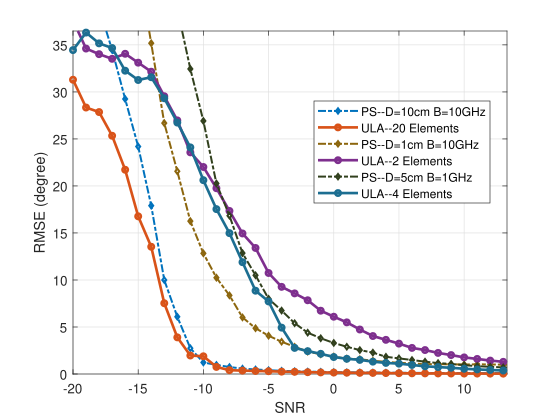}
    \caption{Comparing PS with ULA with half-wavelength gap between array elements in terms of DoA estimation precision. $M=200$, $T_p=100ns$, $B_{rf}=100MHz$} 
    \label{ULAPS}
\end{figure}

In the last simulation we compare the performance of PS technique with  a ULA (beamforming) in terms of DoA estimation precision of a single incoming path. ULA exploits beamforming to steer its beam and compare received power from different angles to find DoA. Figure \ref{ULAPS} depics RSME of DoA estimation for 3 PIAs with different values of $D$ and $B$ and 3 ULAs with different number of array elements. In this simulation, we suppose that ULA is able to integrate the received signal coherently for $T_p$, we also set $T_p=100ns$ and $M=200$, therefore the total integration time of the PIA is $20 \mu s$ . The $B_{rf}$ for ULA and PIA is the same and is set to $100MHz$. As the figure illustrates, the performance of the PIA with $D=10cm$ and $B=10GHz$ is approximately equal to ULA with 20 antennas equally spaced with half wavelength gap (array aperture is 5cm) especially for SNR above -9 dB. Moreover, performance of ULA with 4 elements is close to PIA with $BD=10^8$. Consider that for SNRs above -3 dB, the RMSE is less than $5^o$ for an array with 4 elements, while, beam-width of the array is about $30^o$. If such wide beam antenna uses for communication, the angle estimation precision is much more than what is required. As we discussed in section \ref{DoABW}, angle estimation precision and beamwidth are not coupled and there is no necessity for  antennas of SDMA and IA tasks to be the same. 
Moreover, even when array aperture is small and the number of array elements is few, to obtain a DoA with desirable accuracy a long spatial search is required. For example, to reach an accuracy of $1^o$, any directional antenna with an arbitrary beam-width requires to search at least 180 points to cover a $180^o$ area, in a 2D scenario. On the other hand, to improve PS precision we can simply increase the gap between two antennas and therefore no more complex hardware is required. Furthermore, better precision with ULA requires narrower beams and consequently more time is needed for spatial search to perform the IA task. On the other hand, since no spatial search is required by PS technique, we can obtain an initial guess of DoA very fast, and gradually improve the precision of the estimation by improving SNR through integration.   

\section{Conclusion}
\label{conc}
In this paper, we have introduced DoA estimation via SWR. We have shown that how SWR measures phase difference between two antennas for different frequencies named as PDoAs. We have considered two different implementation schemes for PS: 1- using a long wave-guide to measure amplitude of a standing wave, produced by interaction between two waves received at the two antennas 2- measuring the amplitude of the standing wave inside a short wave-guide for different frequencies of a frequency code-book at different time slots. Moreover, for the second scheme, we have explained that we can use a cascade of multiple PS receivers to measure PDoAs at different frequencies concomitantly. We have developed a signal processing method to extract multiple simultaneous DoAs from PDoAs. We have analyzed processing gain of the technique and discussed that we can take advantage of the required time for spatial search essential for directional techniques to improve DoA estimation precision in PS. Finally, we have analyzed that IA and SD tasks of mobile directional communication can be separated and performed via two dedicated antennas; IA can be performed by PS, SD can be by performed by an array. The separation between these two tasks, reduces delay and overhead and increases communication capacity. Our results have shown that, PS can perform similar to an array, while the required receiver is much less complex than the array receiver, and the spatial search required for DoA estimation can be bypassed.  

 \begin{figure}
    \centering
    \includegraphics[width=4.5in,height=3in]{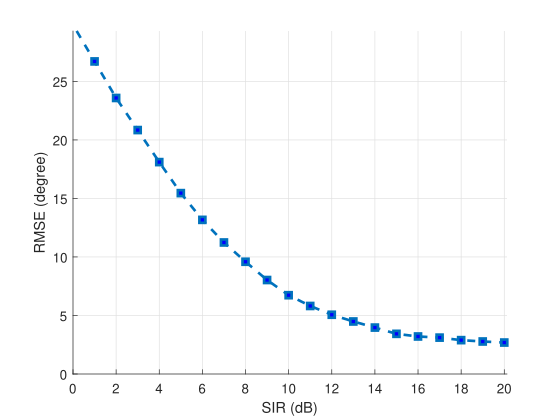}
    \caption{Signal receives at PIA via two paths, one LoS path ($90^o$) and one NLoS path ($30^o$), . $M=40$, $T_p=100ns$, $B_{rf}=100MHz$, $D=1cm$ and source frequency codebook is $[59.5,59.83,60.16,60.5]GHz$.} 
    \label{tightconvarSIR}
\end{figure}

\newpage
\bibliographystyle{IEEEbib}
\bibliography{refs}
\end{document}